\documentclass[fleqn,twoside]{article}
\usepackage{espcrc2}

\usepackage{graphicx}
\usepackage[figuresright]{rotating}

\newcommand{\AmS}{{\protect\the\textfont2
  A\kern-.1667em\lower.5ex\hbox{M}\kern-.125emS}}

\title{Neutral Current $\pi^0$ Interactions at MiniBooNE}

\author{J.L. Raaf\address{University of Cincinnati \\
  Department of Physics, M.L. 0011\\
  Cincinnati, OH 45221, USA} \thanks{\uppercase{T}his work is partially supported by 
grant \uppercase{PHY}-0244915 of the \uppercase{N}ational \uppercase{S}cience 
\uppercase{F}oundation.}~~for the BooNE collaboration \thanks{\uppercase{S}ee 
http://www.boone-fnal.gov/cgi-bin/collaboration/ for entire collaboration list.}}

\begin{document}

\begin{abstract}
MiniBooNE, the Mini Booster Neutrino Experiment at Fermilab, will search 
for the appearance of electron neutrinos in a beam of muon neutrinos, and in the process 
will accumulate more neutrino interactions in the $0-2~\mathrm{GeV}$ energy range than all past experiments. 
This work presents preliminary results from the neutral current $\pi^0$ analysis at MiniBooNE, with
comparisons of the NUANCE, NEUGEN, and NEUT neutrino cross section Monte Carlos. Also presented are
comparisons of these three Monte Carlo simulations with MiniBooNE data taken during the running
period from September, 2002, to September, 2003. 
\vspace{1pc}
\end{abstract}

\maketitle

\section{INTRODUCTION}

\label{sec:intro}
MiniBooNE, the Mini Booster Neutrino Experiment at Fermilab, has been running smoothly and taking data for over a year now. The experiment intends to make a definitive statement about the neutrino oscillation signal seen by the Liquid Scintillator Neutrino Detector (LSND) Experiment at Los Alamos National Laboratory \cite{LSND}. Since both solar and atmospheric neutrino oscillations have been well established in recent years, confirmation of the existence of another oscillation signal with its own distinct mass-squared splitting would indicate a need for further extensions to the Standard Model.

On the way to answering the remaining question of the LSND signal in the neutrino oscillation landscape, MiniBooNE will collect more than 1 million neutrino interactions on pure mineral oil ($\mathrm{CH_2}$). As there is very little experimental input to neutrino interaction cross sections on heavy targets at MiniBooNE energies ($0<\mathrm{E_\nu}<2~\mathrm{GeV}$), the experiment's contributions will be an important addition.

Simulations of neutrino cross sections in the $0-2~\mathrm{GeV}$ region are fairly complex, and are modeled differently by the three cross section Monte Carlos discussed in this work. The MiniBooNE cross sections group has worked in collaboration with the authors of the NUANCE \cite{nuance}, NEUGEN \cite{neugen}, and NEUT \cite{neut} Monte Carlos to provide model comparisons of neutral current $\pi^0$ (NC $\pi^0$) production, comparing the three Monte Carlo simulations with each other and with the MiniBooNE NC $\pi^0$ data set. This is the first time kinematic comparisons of three different neutrino event generators have been made to low energy neutral current $\pi^0$ production data.

\section{MiniBooNE OVERVIEW}
\label{sec:mb_overview}
The Fermilab Booster accelerates protons to $8~\mathrm{GeV}$ kinetic energy in a 15 Hz cycle; 2 to 3 Hz are sent to MiniBooNE. This proton beam is directed into a 71 cm long beryllium target located inside a horn with a toroidal magnetic field. The horn focuses positively charged particles into a 50 meter decay region. Any charged particles that do not decay in the 50 meters are stopped by an absorber located at the downstream end of the decay region. The neutrinos produced in the particle decays then travel through approximately 500 meters of dirt before reaching the detector. The beam at the detector, assuming no oscillations, is very pure -- approximately 99.5\% muon neutrinos with a mean energy of approximately 700 MeV \cite{Monroe}. 

The detector consists of a sphere within a sphere, where the inner signal region is optically isolated from the outer veto shell \cite{BooNEproposal}. The entire detector is instrumented with 8-inch Hamamatsu photomultiplier tubes (PMTs): 1280 in the signal region of the detector (10\% coverage) and 240 in the veto region \cite{PMTtest}. The entire detector is filled with approximately 800 tons of pure mineral oil \cite{Oiltest}.

\begin{figure}[htb]
\begin{center}
\includegraphics[width=2in]{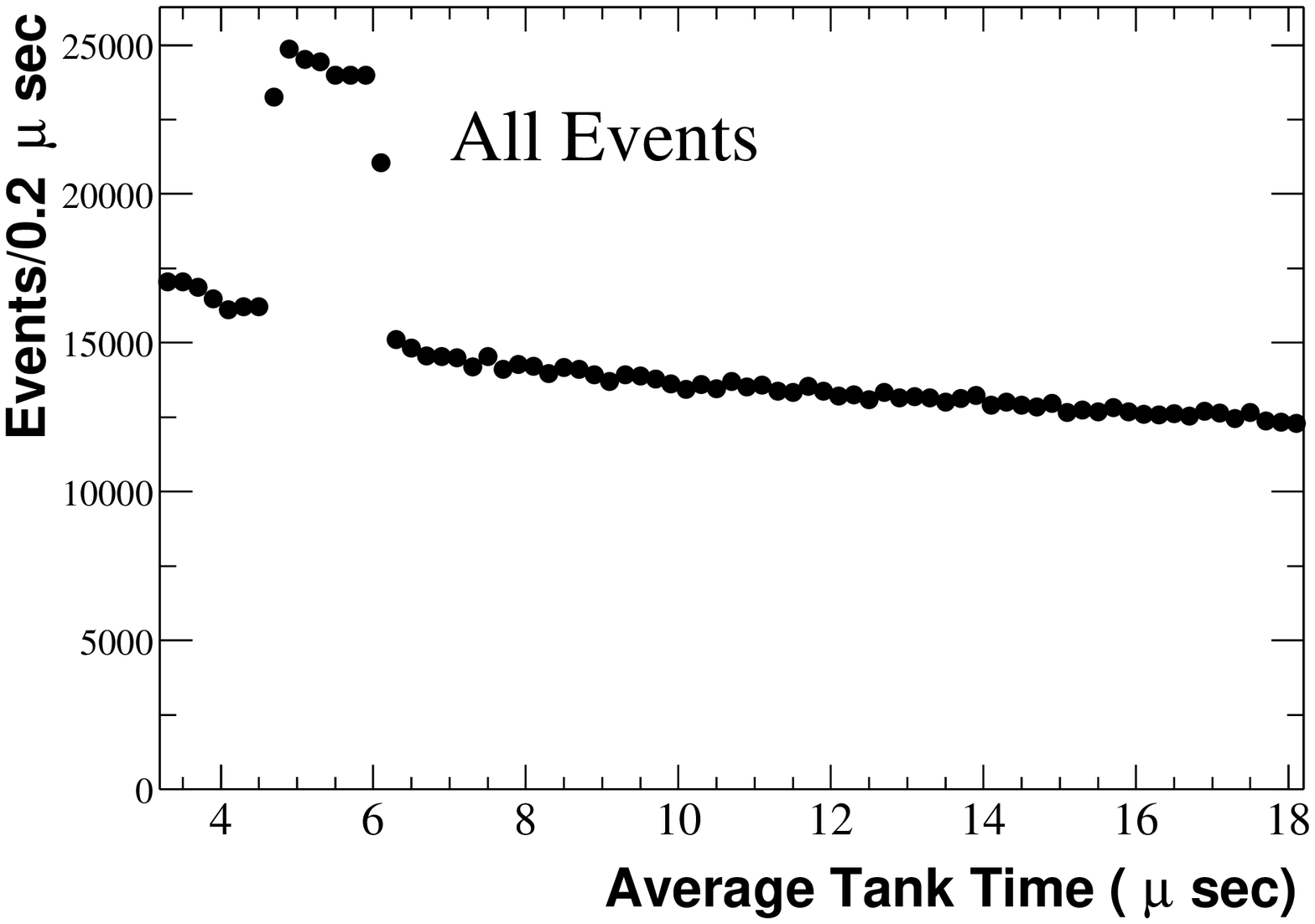}
\includegraphics[width=2in]{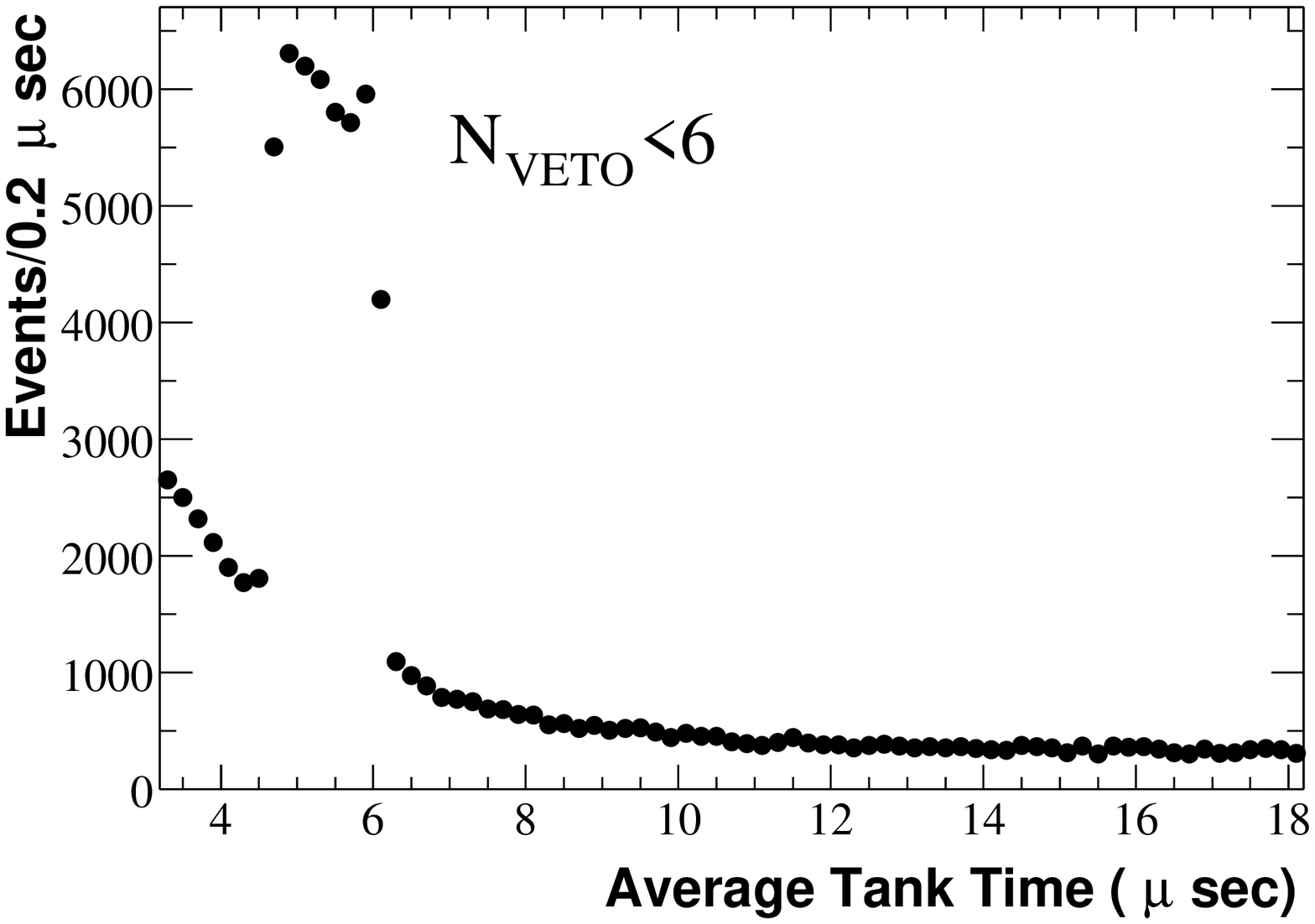}
\includegraphics[width=2in]{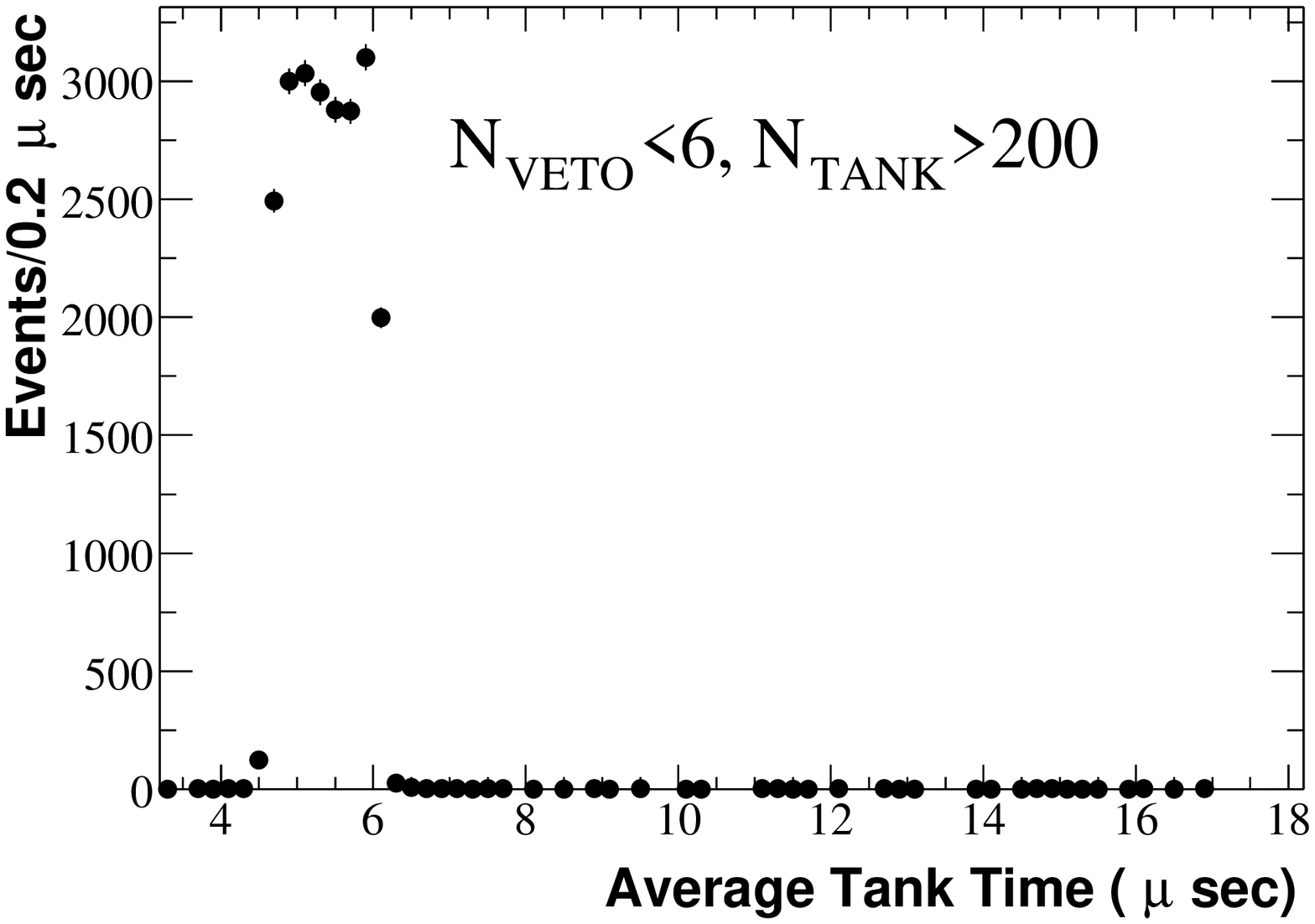}
\end{center}
\vspace{-0.5in}
\caption{The top panel shows recorded tank activity during the beam window with no cuts. The center and bottom panels respectively add a simple veto cut and greater than 200 hits in the main volume of the detector. 
\label{beamtiming}}
\end{figure}

Beam arrives at the target in $1.6~\mu \mathrm{s}$ bursts. The data acquisition system (DAQ) receives a signal from the Booster indicating that beam will be arriving imminently. This signal triggers the DAQ to open a window and record all detector activity for $19.2~\mu \mathrm{s}$, starting $4.8~\mu \mathrm{s}$ before beam arrives. Fig.~\ref{beamtiming} shows the distribution of events during the beam window. The top panel shows the $1.6~\mu \mathrm{s}$ wide peak of the beam particles arriving during the beam window.

With the application of only two very simple cuts, the background on the distribution can be drastically reduced. Requiring no more than 6 hits in the veto region removes events with a coincident cosmic muon occurring in the beam window, shown in the center panel of Fig.~\ref{beamtiming}. The further requirement of at least 200 hits in the main tank region, shown in the bottom panel, largely removes the chance of a muon decaying to an electron during the beam window. With only these two cuts, the signal to background ratio is greater than 1000:1. These cuts provide a very clean sample of beam-induced neutrino interactions to which further cuts may be applied for selection of particular interaction types.

\section{NEUTRAL CURRENT $\pi^0$ PRODUCTION}

\subsection{External Predictions}
\begin{figure}[htb]
\begin{center}
\includegraphics[width=3in]{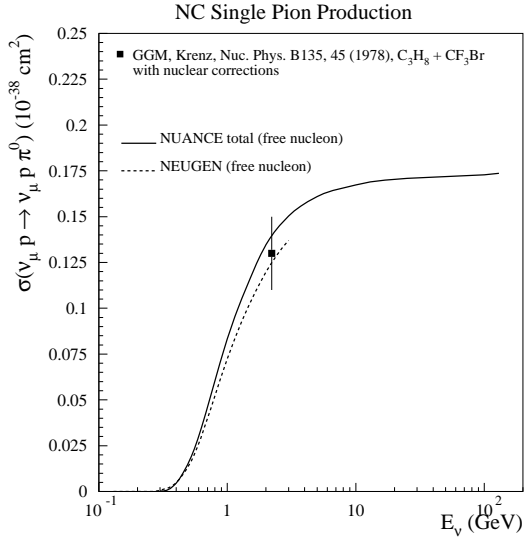}
\end{center}
\vspace{-0.5in}
\caption{Neutral current single $\pi^0$ production cross section on protons as a function of neutrino energy. }
\label{fig:gargamelle}
\end{figure}

There are very few existing measurements of neutral current $\pi^0$ production rates. A re-analysis of 1970's bubble chamber data from Gargamelle \cite{GGM78,EAH_Gargamelle} is shown with cross section predictions from two neutrino cross section Monte Carlos in Fig.~\ref{fig:gargamelle}. The predictions agree with the datum at its energy, but there is currently no confirmation of agreement at other energies. Additional cross section measurements are available as ratios of neutral current to charged current $\pi^0$ production \cite{GGM78,Aachen76,BNL77,BNL79,mauger}, but these measurements can differ by up to factors of 3.

The three neutrino cross section Monte Carlo (MC) simulations compared in this paper -- NUANCE version 3, NEUGEN, and NEUT -- share common theoretical inputs. All use the Llewellyn Smith free nucleon quasi-elastic cross section \cite{LlewellynSmith}, the Rein and Sehgal resonance cross section model \cite{Rein-Sehgal_res}, and the standard deep inelastic scattering (DIS) formula for large momentum transfer \cite{DIS}. There are nontrivial differences among the three though, {\em viz.}, implementation of the Fermi gas model for quasi-elastic interactions, joining of the resonance and DIS regions, and treatment of nuclear effects. The details of these MC simulations may be found in Refs.~\cite{nuance,neugen,neut}, and also in other papers contributed to this set of conference proceedings \cite{Casper,Gallagher,Hayato}. 

Neutral current $\pi^0$'s are created by two main mechanisms: resonant production and coherent production.
Resonant production is dominant, and occurs when a baryon resonance is excited and subsequently decays to a nucleon and one or more mesons, such as $\pi^0$'s. For NC $\pi^0$ resonance production, all three event generators use a Rein and Sehgal-based model. Both NUANCE and NEUT use an axial-vector mass, $\mathrm{m_A}=1.1~\mathrm{GeV/c^{2}}$, while NEUGEN uses $\mathrm{m_A}=1.032~\mathrm{GeV/c^2}$. 

\begin{figure}[htb]
\begin{center}
\includegraphics[width=3in]{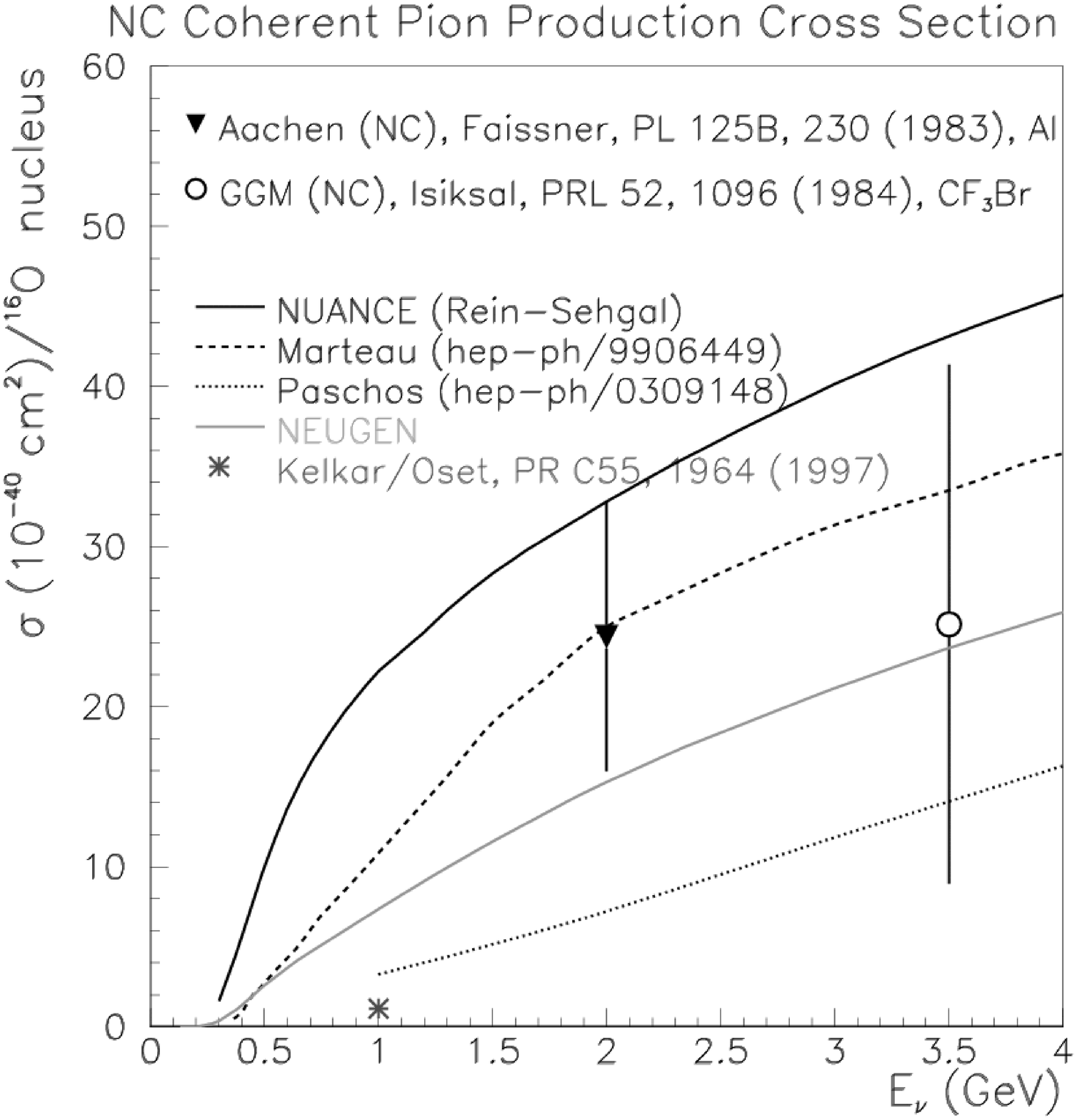}
\end{center}
\vspace{-0.5in}
\caption{Low energy coherent neutral current single $\pi^0$ production as a function of neutrino energy. A recent correction to the NUANCE generator (version 3.004) results in a predicted coherent $\pi^0$ production cross section essentially identical to that shown for NEUGEN.}
\label{fig:coherent_xsec}
\end{figure}

In coherent NC $\pi^0$ production, a neutrino scatters from an entire nucleus rather than the individual constituents of the nucleus. In this process, very little of the momentum is transferred to the recoil nucleus, and the lepton and meson are produced in the forward direction. All three event generators use Rein and Sehgal kinematics and cross sections to describe coherent $\pi^0$ production processes \cite{Rein-Sehgal_coh}. NEUT rescales the Rein and Sehgal cross section down to the level predicted by Marteau \cite{Marteau}, and of the three event generators, NEUGEN predicts the lowest coherent cross section.  The coherent $\pi^0$ cross section as a function of neutrino energy is shown in Fig.~\ref{fig:coherent_xsec} for several models, including NUANCE, NEUGEN, and the Marteau model used to rescale NEUT. It should be noted that the NUANCE prediction shown here does not include pion absorption, which would lower the cross section prediction; absorption effects are included at a later stage in the event generation code. To demonstrate the wide variation in predictions for NC coherent $\pi^0$ production, the Paschos \cite{Paschos} and Kelkar/Oset \cite{Oset} models are also shown. The two data points are from experiments that measured the cross section for coherent $\pi^0$ production at 2 GeV and 3.5 GeV respectively.

\subsection{Analysis Chain}
All Monte Carlo data sets used in this paper were produced by the same method. The MiniBooNE neutrino flux was fed into each of the three cross section Monte Carlo simulations, and the generated events generated were then sent into the full MiniBooNE detector simulation. Events in each sample were reconstructed using the same algorithm. The NC $\pi^0$ events were selected by applying the following cuts to each sample: the background-reducing cuts discussed in section \ref{sec:mb_overview}, a fiducial volume cut (reconstructed event vertex $\mathrm{R}<500~\mathrm{cm}$ from detector center), and a required minimum energy for each of the gammas from $\pi^0$ decay ($\mathrm{E_\gamma}>40~\mathrm{MeV}$).  Using this standard analysis chain for each data set enables us to make a valid comparison of the three MC samples with each other and with data.

Events which pass the event selection cuts have been reconstructed using a $\pi^0$ fitting algorithm that fits events assuming there are two electron-like \v{C}erenkov rings. The event kinematics are determined by the direction and amount of \v{C}erenkov light produced by the gammas from $\pi^0$ decay.

\subsection{Monte Carlo Comparisons}
\label{sec:MCcomparisons}
\begin{figure}[htb]
\begin{center}
\includegraphics[width=3in]{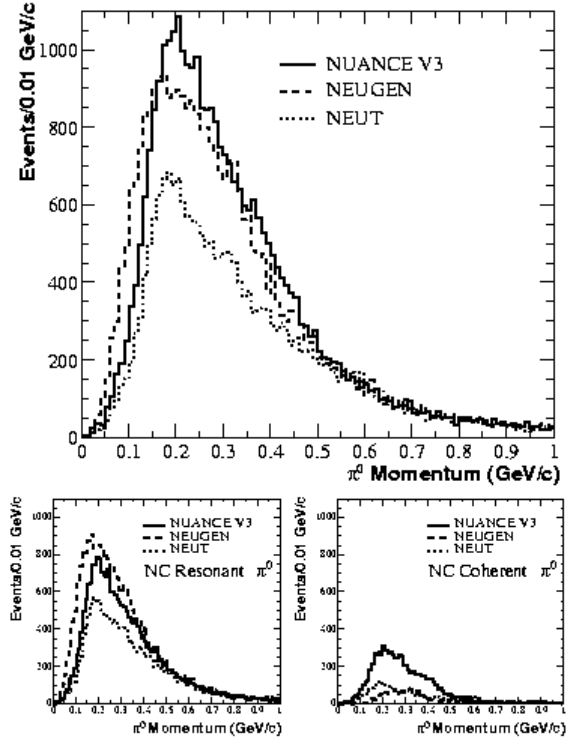}
\end{center}
\vspace{-0.5in}
\caption{$p_{\pi^{0}}$ distribution for Monte Carlo events passing MiniBooNE NC $\pi^0$ event selection cuts: all NC $\pi^0$ (top), only resonant NC $\pi^0$ (bottom left), only coherent NC $\pi^0$ (bottom right).}
\label{fig:compare_ppiMC}
\end{figure}

The momentum distributions for resonant and coherent $\pi^0$'s which pass the NC $\pi^0$ event selection cuts are shown for NUANCE, NEUGEN, and NEUT in the top panel of Fig.~\ref{fig:compare_ppiMC}, normalized to the same number of events before cuts. The overall rate for NC $\pi^0$ production is very similar for NUANCE and NEUGEN, but quite a bit lower for NEUT. We also see that the NEUGEN momentum spectrum is softer than that of both NUANCE and NEUT. The shape difference is more drastic when the distribution is separated into its resonant and coherent components, shown in the bottom left and bottom right panels of Fig.~\ref{fig:compare_ppiMC}, respectively. NEUGEN produces slightly more $\pi^0$'s by resonance excitation than the other two Monte Carlos, and has a harder spectrum for coherent production.  

\begin{figure}[htb]
\begin{center}
\includegraphics[width=3in]{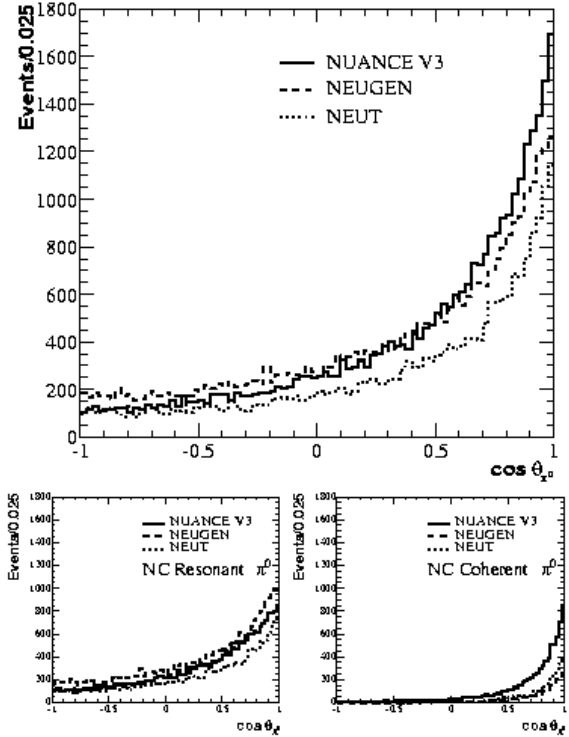}
\end{center}
\vspace{-0.5in}
\caption{$\cos~\theta_{\pi^{0}}$ distribution for Monte Carlo events passing MiniBooNE NC $\pi^0$ event selection cuts: all NC $\pi^0$ (top), only resonant NC $\pi^0$ (bottom left), only coherent NC $\pi^0$ (bottom right).}
\label{fig:compare_ctpiMC}
\end{figure}

Fig.~\ref{fig:compare_ctpiMC} shows the angular distribution of $\pi^0$'s relative to the beam direction for the three MC samples. This distribution is sensitive to the production mechanism -- coherent $\pi^0$'s are much more strongly forward-peaked than resonant $\pi^0$'s, as shown in the bottom panels of this figure. NUANCE produces significantly more coherent $\pi^0$'s than either NEUGEN or NEUT (bottom right panel). The lack of agreement among the three is not surprising, considering the many coherent $\pi^0$ production models that exist and the dearth of data to rule them out.

\subsection{Data/Monte Carlo Comparisons}
\label{sec:dataMCcomparisons}
Understanding the rate and kinematics of $\pi^{0}$ production in neutral current interactions is critical for MiniBooNE, since these events are a major background for the $\nu_{\mu}\rightarrow\nu_{e}$ appearance search. The $\nu_e$ appearance search is a blind analysis; the box will not be opened until all the data have been collected. In the meantime, NC $\pi^0$ events can be used to test Monte Carlo predictions of event kinematics, and will also be used to measure the neutrino neutral current $\pi^0$ cross section. A preliminary analysis of the mass spectrum and kinematics of $\pi^{0}$'s produced in neutral current interactions observed in MiniBooNE is presented here with comparisons to the three Monte Carlo samples discussed above.

\begin{figure}[htb]
\begin{center}
\includegraphics[width=3in]{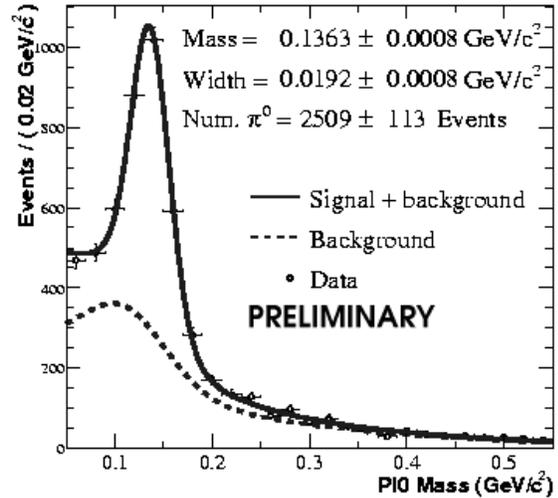}
\end{center}
\vspace{-0.5in}
\caption{Reconstructed invariant mass of beam triggers (open circles with statistical error bars). Fitted shapes are NUANCE-based parameterizations of the contribution from background (dashed curve) and signal (NC resonant and coherent $\pi^{0}$'s, solid minus dashed curve). Errors are statistical only, and do not include systematic errors associated with reconstruction. Note that some of the background events contain $\pi^{0}$'s, so the peak near $m_{\pi^{0}}$ in the dashed curve is expected.}
\label{fig:pi0_mass}
\end{figure}

Fig.~\ref{fig:pi0_mass} shows the reconstructed invariant mass of beam triggers which satisfy the NC $\pi^0$ event selection criteria. Cosmic-ray-induced activity during the beam trigger is eliminated by the method discussed in section \ref{sec:mb_overview}. Further cuts on fiducial volume and energy of $\gamma$'s from $\pi^0$ decay are applied to ensure good reconstruction of the event. The fitted curves are NUANCE-based parameterizations, where the signal contribution arises from NC resonant and coherent single $\pi^{0}$ events, and the background contribution comes from all other events. The background shape is determined by NUANCE version 3 simulations. A background peak near $m_{\pi^{0}}$ is expected, since the background does contain some $\pi^{0}$'s produced in final state interactions and some $\pi^{0}$'s from multi-pion events. For approximately $1\times10^{20}$ protons on target, we extract $2509\pm113$ signal-like NC $\pi^{0}$ events in the fit to data in the preliminary analysis. The fitted mass peak agrees well with the nominal $\pi^{0}$ mass.

The number of neutral current $\pi^{0}$'s seen in data is also extracted in bins of calculated variables of interest ($p_{\pi^{0}}$, $\cos \theta_{CM}$, and $\cos \theta_{\pi^{0}}$) to produce a distribution of each variable with the binned yields. The signal fraction in each bin is extracted via a fit to the invariant mass plot for events in that bin. Unit area normalized distributions for each variable are compared with expectations from NUANCE version 3, NEUGEN, and NEUT.

\begin{figure}[htb]
\begin{center}
\includegraphics[width=3in]{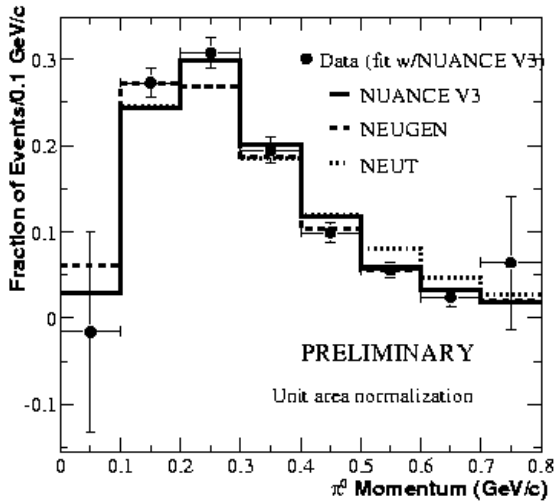}
\end{center}
\vspace{-0.5in}
\caption{Extracted $\pi^{0}$ yields in $p_{\pi^{0}}$ bins comparing Monte Carlo and data, each normalized to unit area. Error bars on data are statistical only.}
\label{fig:ppi}
\end{figure}

The signal $\pi^{0}$ momentum distribution for data is compared to NUANCE, NEUGEN, and NEUT in Fig.~\ref{fig:ppi}. Error bars shown on the data are statistical only. Shape agreement between data and each of the three Monte Carlo samples is shown to be quite good. The fraction of $\pi^{0}$'s reconstructed at high momentum drops off mainly due to the fall-off in the neutrino flux spectrum. 

\begin{figure}[htb]
\begin{center}
\includegraphics[width=3in]{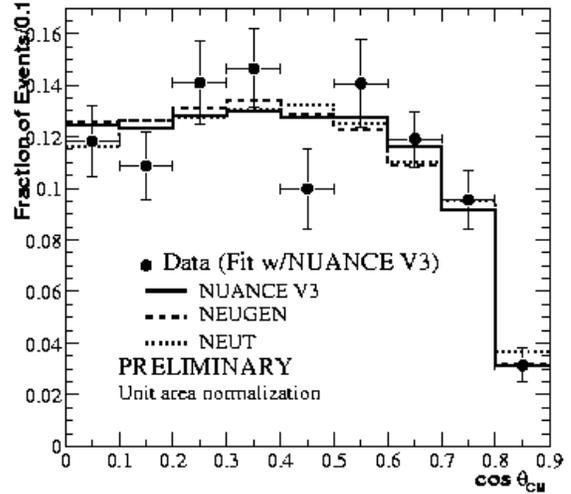}
\end{center}
\vspace{-0.5in}
\caption{Extracted $\pi^{0}$ yields in $\cos~\theta_{CM}$ bins comparing Monte Carlo and data, each normalized to unit area. Error bars on data are statistical only.}
\label{fig:cmpi}
\end{figure}

Fig.~\ref{fig:cmpi} shows the cosine of the $\pi^0$ center of mass angle -- the angle between the direction of the $\pi^0$ in the laboratory frame and the decay axis of the $\gamma$'s in the center of mass frame. This quantity is related to the energy asymmetry of the $\gamma$'s from $\pi^{0}$ decay by $\cos \theta_{\mathrm{CM}} = \frac{1}{\beta}\frac{|{\mathrm{E_1-E_2}}|}{{\mathrm{E_1+E_2}}}$. Because the $\pi^0$ is pseudoscalar, it has no preferred decay direction and the center of mass angular distribution should be flat. The spectrum is distorted, however, for $\cos\theta_{\mathrm{CM}}>\sim0.7$ because of the minimum energy requirement for each of the $\gamma$'s in the decay; events where one or both of the $\gamma$'s have less than 40 MeV are cut. All three Monte Carlo samples model the data well. This is especially important in the region where the spectrum falls off because these are the class of events that will mimic an oscillation signal.

\begin{figure}[htb]
\begin{center}
\includegraphics[width=3in]{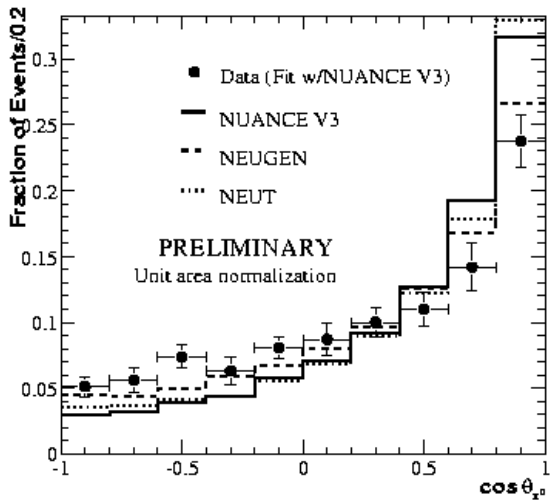}
\end{center}
\vspace{-0.5in}
\caption{Extracted $\pi^{0}$ yields in $\cos~\theta_{\pi^{0}}$ bins comparing Monte Carlo and data, each normalized to unit area. Error bars on data are statistical only.}
\label{fig:ctpi}
\end{figure}

The angular distribution of $\pi^{0}$'s relative to the beam direction is shown in Fig.~\ref{fig:ctpi}. As stated in section~\ref{sec:MCcomparisons}, this distribution is sensitive to the $\pi^0$ production mechanism; it may help us to determine how much coherent production contributes to the overall rate of NC $\pi^{0}$ production. The data appear to have a much lower contribution from coherent $\pi^0$'s than all three Monte Carlos predict, as indicated by the deficit in the forward-most bin. 

\section{CONCLUSIONS}
The work presented here is the first time several low energy Monte Carlo event generators in use by various experiments have been uniformly compared to experimental distributions of NC $\pi^0$ events. These preliminary comparisons to the MiniBooNE data are both interesting and encouraging. The shapes of kinematic distributions are surprisingly similar despite their differing theoretical inputs. The exception is in the $\pi^0$ angular distribution relative to the beam direction, where there is an expected difference due to the different coherent contributions among the three Monte Carlo simulations. The deficit in the forward-most bin of Fig.~\ref{fig:ctpi} seems to indicate a much lower contribution from coherent $\pi^0$ production than any of the MC simulations predict; more data are needed, however,  before any conclusions may be drawn. There is a serious lack of NC $\pi^0$ cross section data at low energies for both resonant and coherent NC $\pi^0$ production; MiniBooNE's complete collection of NC $\pi^0$ data, which will ultimately be $\sim10$ times the data presented here, will provide an important measure of this cross section in the $0-2~\mathrm{GeV}$ region.

\section{ACKNOWLEDGEMENTS}
The author would like to thank D. Casper, H. Gallagher, and Y. Hayato for providing Monte Carlo samples generated with the MiniBooNE flux, and also for useful discussions about their programs. Many thanks also go to INFN, who generously provided travel support. The financial support of both the NSF and the DOE are gratefully acknowledged.

\end{document}